# Exact and approximate solutions for a century-old problem: A general treatment of Henri-Michaelis-Menten enzyme kinetics

Mário N. Berberan-Santos[a]

Centro de Química-Física Molecular and IN - Institute of Nanoscience and Nanotechnology, Instituto Superior Técnico, Universidade Técnica de Lisboa, 1049-001 Lisboa, Portugal

[a]Electronic mail: berberan@ist.utl.pt.

**ABSTRACT**

A different view of Henri-Michaelis-Menten (HMM) enzyme kinetics is presented. In the first part of the paper, a simplified but useful description that stresses the cyclic nature of the catalytic process is introduced. The time-dependence of the substrate concentration after the initial transient phase is derived in a simple way that dispenses the mathematical technique known as quasi-steady-state approximation. In the second part of the paper an exact one-dimensional formulation of HMM kinetics is obtained. The whole problem is condensed in a single one-variable evolution equation that is a second-order non-linear differential equation, and the control parameters are reduced to three dimensionless quantities: enzyme efficiency, substrate reduced initial concentration, and enzyme reduced initial concentration. The exact solution of HMM kinetics is obtained as a set of Maclaurin series. From the same equation, a number of approximate solutions, some known, some new, are derived in a systematic way that allows a precise evaluation of the respective level of approximation and conditions of validity. The evolution equation derived is also shown to be well suited for the numerical computation of the concentrations of all species as a function of time for any given combination of parameters.

# I. INTRODUCTION

Enzyme *E* and substrate *S* (usually in large excess) associate by a fast bimolecular reaction ($k_a \approx 10^7$ to $10^{10}$ M$^{-1}$ s$^{-1}$) to form the enzyme-substrate complex *ES*. The substrate is then transformed into product *P* and the complex finally dissociates releasing the product and regenerating the enzyme. However, the reaction is in general reversible, and the complex can also dissociate back to give free substrate and free enzyme. All these elementary steps are condensed in the well-known Henri-Michaelis-Menten (HMM) kinetic scheme,[1,2] which is the simplest description of enzyme action:

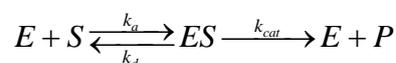

**Scheme 1**

In this scheme, reversibility of the second step is neglected, and also the necessary existence of a second complex *EP* is not accounted for explicitly. With respect to the more general Haldane scheme,[3] the HMM description holds whenever the ES/EP interconversion is relatively fast and the reversibility in the last step can be neglected. This will be assumed throughout.

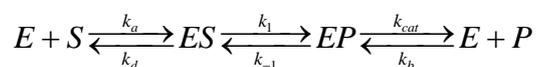

**Scheme 2**

An important quantity defining enzyme activity is the turnover rate $r_t$,[1,2]

$$r_t = \frac{1}{[E]_0} \frac{d[P]}{dt} \tag{1}$$

where $[E]_0 = [E]+[ES]$. The turnover rate is the number of product molecules generated per unit time and per enzyme molecule present. This quantity is especially meaningful when the enzyme concentration is much lower than that of substrate, which is the most common situation.

## II. THE BASIC EQUATIONS

The differential equations of the HMM mechanism relating the concentrations of the four species present in a batch reactor are:

$$\frac{d[S]}{dt} = -k_a[E][S] + k_d[ES], \qquad (2)$$

$$\frac{d[ES]}{dt} = k_a[E][S] - (k_{cat} + k_d)[ES], \qquad (3)$$

$$\frac{d[E]}{dt} = -k_a[E][S] + (k_{cat} + k_d)[ES], \qquad (4)$$

$$\frac{d[P]}{dt} = k_{cat}[ES]. \qquad (5)$$

These equations embody the two mass conservation equations

$$[E]_0 = [E] + [ES], \qquad (6)$$

$$[S]_0 = [S] + [ES] + [P]. \qquad (7)$$

The state of the system is thus defined by the concentrations of any two species. If the mass conservation equations are used, only two differential equations are needed.

Suitable additional assumptions for a batch reactor are $[ES]_0 = 0$ and $[P]_0 = 0$. The kinetics is thus controlled by five parameters: Two initial concentrations, $[S]_0$, $[E]_0$, and three rate constants, $k_a$, $k_d$, and $k_{cat}$.

Explicit relations can be obtained between the concentrations. Using Eqs. (5)-(7) the concentrations of $S$, $E$ and $ES$ can all be written in terms of the concentration of $P$:

$$[S] = [S]_0 - [P] - \frac{1}{k_{cat}}\frac{d[P]}{dt}, \qquad (8)$$

$$[ES] = \frac{1}{k_{cat}}\frac{d[P]}{dt}, \qquad (9)$$

$$[E] = [E]_0 - \frac{1}{k_{cat}}\frac{d[P]}{dt}. \qquad (10)$$

This means that if an explicit form for $[P](t)$ is obtained, all other three concentrations are also easily computed.

Eqs. (8)-(10) can also be reversed in order to obtain the concentration of $P$ in terms of the concentrations of $S$, $E$ or $ES$. In particular,

$$[P] = [S]_0 \left(1 - e^{-k_{cat}t}\right) - k_{cat}[S] \otimes e^{-k_{cat}t}, \tag{11}$$

where $\otimes$ stands for the convolution between two functions, $f \otimes g = \int_0^t f(u)g(t-u)du$. Eq. (11) is obtained from Eq. (8) by taking Laplace transforms, solving the equation for the Laplace transform of $[P]$, and finally inverting the transforms. In this way, computation of $[P]$ from $[S]$ requires the past history of $[S]$ to be known, while computation of $[S]$ from $[P]$ requires only the present behavior of $[P]$. This asymmetry stems from the role of these species, as $S$ is a precursor of $P$, and there is no feedback on $S$ from $P$.

No general solution was up to now known for the HMM kinetics, that is, the system of two first-order differential equations had not been solved to yield expressions for the concentrations of all species neither in closed form (i.e., in terms of known functions) nor in terms of power series expansions.

## III. QUASI-STEADY-STATE APPROXIMATION

The traditional approach to HMM kinetics, pioneered by Briggs and Haldane,[1,2] has been to reduce the system of two first-order differential equations to a single first-order differential equation by using the quasi-steady-state approximation (QSSA), according to which one of the differential equations, Eq. (3) or Eq. (4), is converted into an algebraic equation, by equating the time derivative to zero. In this treatment, based on observations corresponding to an experimentally meaningful but limited region of parameter space, it is assumed that after a short period (transient kinetics) the concentration of enzyme-substrate complex $ES$ is approximately constant, and that this implies a time derivative of $[ES]$ that is precisely zero. This somewhat drastic mathematical procedure converts Eqs. (3) and (4) in a simple algebraic equation, and the turnover rate becomes[1,2]

$$r_t = \frac{k_a k_{cat}[S]}{k_d + k_{cat} + k_a[S]} = \frac{k_{cat}[S]}{K_m + [S]}, \tag{12}$$

where $K_m$ is the Michaelis constant,

$$K_m = \frac{k_d + k_{cat}}{k_a}. \tag{13}$$

Within the QSSA approximation, this constant defines the fraction of bound enzyme for a given substrate concentration

$$\frac{[ES]}{[E]+[ES]} = \frac{[S]}{K_m + [S]}. \tag{14}$$

If $[S] \gg K_m$, then practically all enzyme exists in bound form, saturation is reached, and the turnover rate attains its maximum value, $k_{cat}$, also called the turnover number. For $[S] = K_m$ the turnover rate is half of the maximal value. Typical turnover numbers are 100 to 1000 s$^{-1}$, with maximum values of about 10$^6$ s$^{-1}$. Note that $\frac{k_{cat}}{K_m} = \frac{k_a}{1 + \frac{k_d}{k_{cat}}} < k_a$ and therefore this ratio gives a minimum value estimate for $k_a$.

The concentration of enzyme-substrate complex is in fact never constant, although it may change very slowly. The QSSA mathematical approximation only holds exactly at a precise time, when the *ES* concentration attains its maximum value, and the initial arguments invoked for its use were qualitative. Only relatively recently were the limits of applicability of the QSSA to HMM kinetics thoroughly investigated.[4-11]

# IV. A DIFFERENT VIEW OF ENZYME KINETICS

## IV.1 Catalytic cycles

Enzyme activity can be advantageously depicted as a simple catalytic cycle,

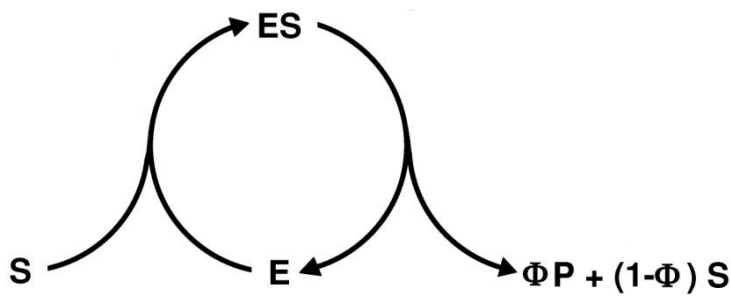

**Scheme 3**

Reversibility is incorporated in the diagram by assigning an effective stoichiometric coefficient $\Phi$ to the product. This coefficient, which takes values between 0 and 1, is the enzyme efficiency i.e. the product yield per cycle, given by[12]

$$\Phi = \frac{k_{cat}}{k_d + k_{cat}} = \frac{k_{cat}}{k_a K_m}. \tag{15}$$

Super-efficient enzymes,[13] for which $k_d \ll k_{cat}$, have a $\Phi$ close to unity, and for $[S] \ll K_m$ the reaction for these enzymes can be nearly diffusion-controlled, as $r_t = k_a [S]$. When $\Phi \ll 1$ the kinetic situation corresponds to a pre-equilibrium between reactants and enzyme-substrate complex, and $K_m$ reduces to the equilibrium constant for the dissociation reaction. This is the case originally considered by Michaelis and Menten.[1,2]

The inverse of $\Phi$, $\bar{n}_P$, is the average number of cycles required for conversion of one substrate molecule into one product molecule,

$$\bar{n}_P = \frac{1}{\Phi} = 1 + \frac{k_d}{k_{cat}}. \tag{16}$$

The product yield per cycle is an important parameter for characterizing enzyme activity, but it does not tell at which frequency catalytic cycles proceed. What is thus the duration (or period) of a catalytic cycle? This duration is not strictly the same for all cycles, but has an average value $\tau_c$ given by[12]

$$\tau_c = \frac{1}{k_a[S]} + \frac{1}{k_d + k_{cat}} = \frac{1}{k_a}\left(\frac{1}{[S]} + \frac{1}{K_m}\right). \tag{17}$$

The cycle duration has two terms, corresponding to the two halves of the catalytic cycle: The first term, $1/(k_a[S])$, is the average duration (or lifetime) of the enzyme in the free state. This duration is inversely proportional to the free substrate concentration. The second term, $1/(k_d+k_{cat})$, is the average duration (or lifetime) of the enzyme-substrate complex, and is independent of the concentration of substrate.

Using Eqs. (15) and (17) the turnover rate can be rewritten in a simple way,[12]

$$r_t = \frac{\Phi}{\tau_c}. \tag{18}$$

The meaning of Eq. (18) is clear: The turnover rate is the number of successful $E \to ES \to E$ cycles per unit time and per enzyme molecule. Increase of the substrate concentration does not change the yield of product formation per cycle $\Phi$, but reduces the average cycle duration $\tau_c$, as the free enzyme lasts less time before forming the complex with the substrate. However, above a certain substrate concentration (saturation conditions), the cycle duration attains a constant, minimum value, $1/(k_d+k_{cat})$, controlled by the complex lifetime, and therefore the turnover rate reaches its maximal value $k_{cat}$.

For a reaction in a closed vessel (batch reactor) it may be asked: In how many catalytic cycles is each enzyme (on the average) engaged during the entire course of the reaction? To answer this question, we first note that $[S]_0/[E]_0$ substrate molecules are converted into product molecules by each enzyme. In this way, and using Eq. (16), the total average number of cycles per enzyme for the complete course of the reaction is

$$\bar{n}_c = \frac{\bar{n}_P [S]_0}{[E]_0}. \tag{19}$$

If we now consider the situation at time *t*, then the average number of cycles per enzyme up to this point is

$$\bar{n}_c(t) = \frac{\bar{n}_P [\mathrm{P}](t)}{[\mathrm{E}]_0}. \tag{20}$$

For efficient enzymes that are in large excess, one cycle suffices for the reaction to attain completion. In fact, all complexes form almost simultaneously, the cycles are completed at approximately the same time, and the reaction is terminated in one period. If the enzymes are less efficient, several periods are required, but all substrate molecules have a common start and react in a parallel, noncompetitive fashion. The situation $[E]_0 \gg [S]_0$ has therefore little practical interest, as the reaction would be very fast, and even rapid mixing techniques could not be used, except for extremely low enzyme concentrations.

When it is the substrate that is in excess, however, most substrate molecules must wait for a chance to form a complex with an available enzyme. The reaction takes much longer and needs a different description.

**IV.2 Substrate in excess**

The case where $[S]_0 \gg [E]_0$ is the more common situation. Typically, enzyme concentrations of $10^{-8}$ to $10^{-10}$ M are used, and substrate concentrations are usually greater than $10^{-6}$ M. Under these conditions the enzymes, even when super-efficient, must perform many catalytic cycles. Eqs. (1), (17), and (18) give

$$\frac{d[P]}{dt} = \frac{k_a \Phi [E]_0}{\dfrac{1}{[S]} + \dfrac{1}{K_m}} = \frac{k_{cat}[E]_0}{1 + \dfrac{K_m}{[S]}}. \tag{21}$$

Eq. (21) is valid if many catalytic cycles take place for each enzyme before completion of the reaction that is, if substrate concentration changes slowly compared to the cycle duration. Using Eq. (19), this implies that $[S]_0 \gg [E]_0 \Phi$, a condition that clearly holds if the total enzyme concentration $[E]_0$ is much less than the substrate concentration $[S]_0$. In order to proceed, it is noted that $[S]_0 \gg [E]_0$ implies that $[S] \simeq [S]_0 - [P]$, and using this relation Eq. (21) can be integrated in order to obtain $[P](t)$. The concentration of substrate is next obtained from the same equation $[S] \simeq [S]_0 - [P]$. The final result is well-known in both implicit[14] and explicit[7] forms. The explicit form is[7]

$$[S](t) = K_m W\left[\left(\frac{[S]_0}{K_m}\right)\exp\left(\frac{[S]_0}{K_m}\right)\exp\left(-\frac{k_{cat}[E]_0}{K_m}t\right)\right], \qquad (22)$$

where $W(x)$ is the Lambert function, see Appendix A. A more general solution is given in Sect. VI.

A second non-standard derivation also leading to Eq. (22) starts from Eq. (7). Differentiating it with respect to time, and taking into account that during the so-called quasi-steady-state, the concentration of *ES* changes much less rapidly than that of *S*,

$$\frac{d[P]}{dt} \simeq -\frac{d[S]}{dt}. \qquad (23)$$

Insertion of Eq. (23) into Eq. (21) gives

$$-\frac{d[S]}{dt} = \frac{k_{cat}[E]_0}{1 + \frac{K_m}{[S]}}. \qquad (24)$$

This equation is identical to that obtained from the QSSA.[1,2] The present derivations differ however from that based on the QSSA and are much more satisfactory. The drastic mathematical procedure followed when applying the QSSA (i.e. equating the derivative of [*ES*] with respect to time to zero) is not necessary for arriving at Eq. (24) or to the integrated solution (see Appendix A). According to the first derivation, there is even no need to consider the time derivative of [*ES*], it being enough to take into account that when $[S]_0 \gg [E]_0$ the condition $[ES] \ll [S]+[P]$ is obeyed for all times.

It is interesting at this point to mention two limiting situations:

(i) If $[S]_0 \gg K_m$, the substrate concentration obeys zero-order kinetics (down to $[S] \approx K_m$):

$$[S] = [S]_0 - k_{cat}[E]_0 t. \tag{25}$$

Historically, the observation of this behavior provided striking kinetic evidence for the existence in solution of a relatively stable enzyme-substrate complex, entity whose existence had already been postulated by Emil Fischer and others.

(ii) If $[S]_0 \ll K_m$ (but still with $[S]_0 \gg [E]_0$), the substrate concentration obeys first-order kinetics

$$[S] = [S]_0 \exp\left(-k_a \Phi [E]_0 t\right) = [S]_0 \exp\left(-\frac{k_{cat}[E]_0}{K_m} t\right), \tag{26}$$

The corresponding equations for the product are obtained from $[P] = [S]_0 - [S]$.

Eqs. (22), (25) and (26) show that by changing the substrate initial concentration, and by following the time course of the reaction, it is possible to determine the values of $k_{cat}$ and $K_m$. This is accomplished by a number of data analysis methods.[2,9,15]

## IV.3 Pre-equilibrium

Under pre-equilibrium conditions[16] $\Phi \ll 1$, a fast equilibrium between $S$, $E$ and $ES$ is rapidly established, and the conversion of $ES$ into $P$ takes place much more slowly. When $[S]_0 \gg [E]_0$ the results of the previous section apply. Nevertheless, Eq. (21) is still valid when the substrate is not in excess, provided that $[S]_0 \gg [E]_0 \Phi$. In such a case the $ES$ concentration is significant and $[S] \simeq [S]_0 - [P]$ is not valid. It is then preferable to write $[S'] = [S]_0 - [P]$, where $[S']$ is the total substrate concentration, $[S'] = [S] + [ES]$, and to change the variable in Eq. (21) to $[S']$. Using the equilibrium constant, $[S]$ can then be related to $[S']$ and the equation finally integrated. The detailed procedure will be presented in Sect. VI.

## V. THE INITIAL TRANSIENT PHASE

The full set of three rate constants ($k_a$, $k_d$, and $k_{cat}$) can only be obtained from an additional study of the initial time-dependence of the concentration during the so-called transient phase,[17] the initial period that lasts less than the average cycle time τ$_c$ (a few *ms* at most). During at least the early part of this phase, the substrate concentration is approximately equal to its initial value, and therefore the situation is well represented by Scheme 4,

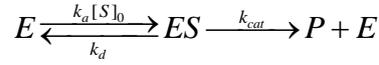

**Scheme 4**

The concentrations of free enzyme and enzyme-substrate complex are

$$[E](t) = [E]_0 \frac{K_m + [S]_0 \exp\left[-k_a\left(K_m + [S]_0\right)t\right]}{K_m + [S]_0} = [E]_0 \left(1 - k_a[S]_0 t\right) + ..., \quad (27)$$

$$[ES](t) = k_a[E]_0[S]_0 \frac{1 - \exp\left[-k_a\left(K_m + [S]_0\right)t\right]}{K_m + [S]_0} = k_a[E]_0[S]_0 t + .... \quad (28)$$

For very short times the free enzyme and the enzyme-substrate complex concentrations evolve linearly with time, and for longer times they stabilize at the steady-state values.

The concentration of product follows from Eq. (5) and (28),[17,18]

$$[P](t) = \frac{k_{cat}[E]_0[S]_0}{K_m + [S]_0}\left(t - \frac{1 - \exp\left[-k_a\left(K_m + [S]_0\right)t\right]}{K_m + [S]_0}\right) = \frac{1}{2}\frac{k_{cat}[E]_0[S]_0}{K_m + [S]_0} t^2 + ... \quad (29)$$

For very short times the product evolves in a quadratic way with time, and for longer times it increases linearly with time. For still longer times it stabilizes and attains a constant value $[S]_0$ but owing to the assumptions made this final phase is not predicted by Eq. (29).

The overall rate constant obtainable from Eqs. (27)-(29) is $k_d + k_{cat} + k_a[S]_0$ and from a study of its dependence with $[S]_0$ the bimolecular rate constant $k_a$ can be obtained, along with $k_d + k_{cat}$.

# VI. A UNIFIED APPROACH

## VI.1 A single evolution equation and its exact solution

All available treatments of HMM kinetics start from the exact system of two first-order non-linear differential equations. Each approximate treatment results from specific simplifications made on this system. As discussed, the QSSA and pre-equilibrium approximations lead to a single first-order non-linear differential equation, while the case of excess enzyme leads to two coupled first-order and linear differential equations. The discussion of the kinetics in terms of analytical and numerical integration methods (e.g. perturbation theory) is also commonly carried out in a two-dimensional phase space.[4-6, 10, 11, 19]

An alternative approach to HMM kinetics will now be presented. As is well known, a single second-order differential equation can be split into an equivalent system of two first-order differential equations. The inverse procedure, although less common, is also possible. It will be shown here that it can be advantageously used in the study of HMM kinetics.

Differentiation of Eq. (5) with respect to time, followed by the expression of all concentrations in terms of [P], Eqs. (8)-(10), gives

$$\frac{d^2[P]}{dt^2} = -\left[k_a\left([E]_0 + [S]_0 - [P]\right) + k_d + k_{cat} - \frac{k_a}{k_{cat}}\frac{d[P]}{dt}\right]\frac{d[P]}{dt} + k_a k_{cat}[E]_0\left([S]_0 - [P]\right), \quad (30)$$

with the initial conditions $[P](0) = [P]'(0) = 0$. This equation appears unwieldy, but when rewritten in terms of the dimensionless variables $e = [E]/K_m$, $s = [S]/K_m$, $c = [ES]/K_m$, $p = [P]/K_m$, $\tau = k_{cat} t$, and the dimensionless total substrate concentration $\sigma = s+c = s_0-p$, and with $e_0 = [E]_0/K_m$ and $s_0 = [S]_0/K_m$, it becomes considerably simpler:

$$\Phi\frac{d^2\sigma}{d\tau^2} + \left(1 + e_0 + \sigma + \frac{d\sigma}{d\tau}\right)\frac{d\sigma}{d\tau} + e_0\sigma = 0, \quad (31)$$

where $\Phi$ is the enzyme efficiency, Eq. (15). The initial conditions are $\sigma(0) = s_0$ and $\sigma'(0) = 0$. The variable $\sigma$ is a positive and monotonically decreasing function of the reduced time $\tau$, and $\sigma(\infty) = 0$. As mentioned, $\sigma$ is the reduced substrate concentration, present both as free substrate and as part of the enzyme-substrate complex. The dimensionless parameters defined follow a natural choice for the simplification of the dimensional equation, have a clear physical meaning, and are homogeneous in the sense that all dimensionless concentrations are defined in the same way. The whole problem is condensed in a single one-variable evolution equation, which is a second-order non-linear differential equation, and the control parameters are reduced to just three dimensionless quantities: the enzyme efficiency $\Phi$, the substrate reduced initial concentration, $s_0$, and the enzyme reduced initial concentration, $e_0$. The reduced variables absorb the other two parameters.

Eq. (31) is an exact one-dimensional formulation of HMM kinetics. From it, one can derive in a systematic way a number of approximate solutions, as will be shown. It is also quite useful for the computation of the concentrations of all species as a function of time for any given combination of parameters ($\Phi$, $e_0$, $s_0$). In fact, once obtained $\sigma(\tau)$,

$$s = \sigma + \frac{d\sigma}{d\tau}, \tag{32}$$

$$e = e_0 + \frac{d\sigma}{d\tau}, \tag{33}$$

$$c = -\frac{d\sigma}{d\tau}, \tag{34}$$

$$p = s_0 - \sigma. \tag{35}$$

Numerical integration of Eq. (31) is routinely carried out with technical software such as *Mathematica*, and dispenses consideration of special numerical methods.

Furthermore, the exact solution of HMM kinetics can be obtained as a set of Maclaurin series by rewriting Eq. (31) as

$$\frac{d^2\sigma}{d\tau^2} = -\frac{1}{\Phi}\left(1 + e_0 + \sigma + \frac{d\sigma}{d\tau}\right)\frac{d\sigma}{d\tau} - e_0\sigma, \tag{36}$$

and computing successive derivatives at the origin. In this way, and using also $\sigma(0) = s_0$ and $\sigma'(0) = 0$,

$$\sigma = s_0 - \frac{e_0 s_0}{\Phi}\frac{\tau^2}{2} + \frac{e_0 s_0}{\Phi^2}(1 + e_0 + s_0)\frac{\tau^3}{3!} - \frac{e_0 s_0}{\Phi^3}\left[2e_0 s_0 + (1 + e_0 + s_0)^2 - \Phi e_0\right]\frac{\tau^4}{4!} + \dots, \tag{37}$$

and from Eqs. (32), (34) and (35),

$$s = s_0 - \frac{e_0 s_0}{\Phi}\tau + \frac{e_0 s_0}{\Phi^2}(1 - \Phi + e_0 + s_0)\frac{\tau^2}{2} + \dots, \tag{38}$$

$$c = \frac{e_0 s_0}{\Phi}\tau - \frac{e_0 s_0}{\Phi^2}(1 + e_0 + s_0)\frac{\tau^2}{2} + \dots, \tag{39}$$

$$p = \frac{e_0 s_0}{\Phi}\frac{\tau^2}{2} - \frac{e_0 s_0}{\Phi^2}(1 + e_0 + s_0)\frac{\tau^3}{3!} + \dots. \tag{40}$$

No attempt has been made to obtain the general forms of the terms of these series, but the number of coefficients that can be obtained with a software like *Mathematica* is in principle virtually unlimited, especially numerically.

It may be remarked that in order to accommodate $c_0 > 0$ in the above computational scheme the only required modification is to take into account that $\sigma'(0) = -c_0$.

As far as the author is aware, the general HMM kinetics one-variable second-order differential equation was not derived before in any form (dimensional or dimensionless). Only a simplified form (valid for low total enzyme concentration)[17,18] was used and applied without special advantage to the transient phase, whose resulting equation is linear (see also below Sect. VI.2.3).

## VI.2 Approximate solutions

### VI.2.1 Quasi-steady-state and pre-equilibrium approximations

As written, Eq. (31) has three terms. The first, $\Phi \dfrac{d^2\sigma}{d\tau^2}$, can be recast as

$$\Phi\frac{d^2\sigma}{d\tau^2} = -\Phi\frac{dc}{d\tau} = c - e\,s. \tag{41}$$

It starts from a negative value, $-e_0 s_0$, increases and crosses zero, becomes positive, passes through a maximum and then decreases towards zero, being relatively small and positive after the maximum, see Fig. 1.

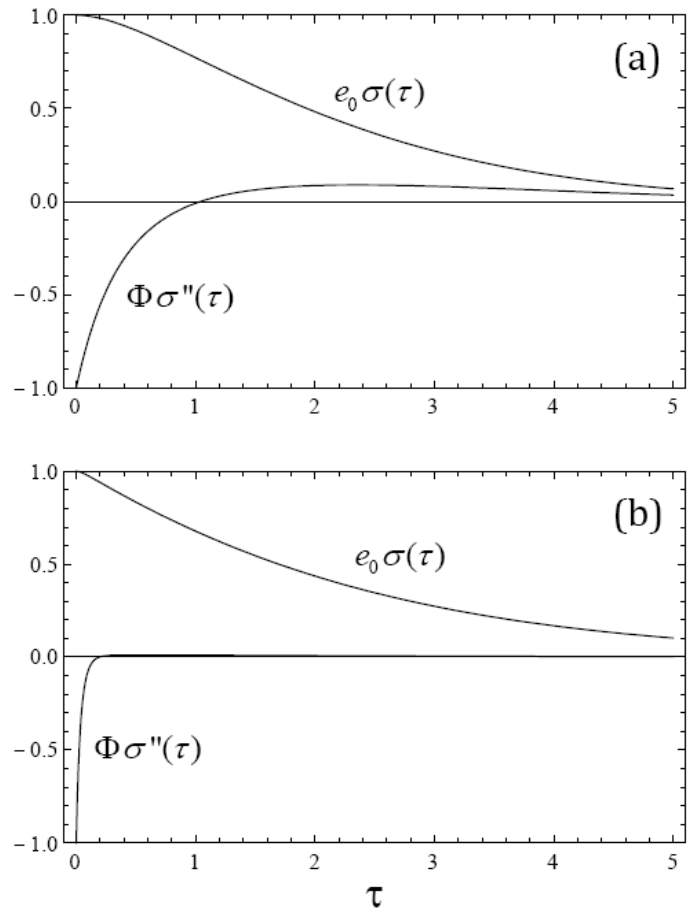

**Fig.1** Plot of $\Phi\dfrac{d^2\sigma}{d\tau^2}$ and of $e_0\sigma$ as a function of the reduced time $\tau$ for $e_0 = s_0 = \Phi = 1$ (a), and $e_0 = s_0 = 1$, and $\Phi = 0.1$ (b). When $\Phi \ll 1$ (pre-equilibrium) the transient term $\Phi\dfrac{d^2\sigma}{d\tau^2}$ is negligible except for very short times.

The second term, $\left(1+e_0+\sigma+\dfrac{d\sigma}{d\tau}\right)\dfrac{d\sigma}{d\tau}$, is always negative, as the bracketed quantity that can be rewritten as $1+e_0+s$ (cf. Eq. (32)), is always positive, and $\sigma$ has a negative time derivative for all times. Finally, the third term, $e_0\sigma$, is always positive.

As the three terms must add up to zero, the first one is negligible in face of the other two if these are of identical magnitude (but with opposite signs), hence the first term can be dropped if

$$\Phi\left|\dfrac{d^2\sigma}{d\tau^2}\right| \ll e_0\sigma, \tag{42}$$

or

$$\Phi\left|\dfrac{dc}{d\tau}\right| \ll e_0(s+c), \tag{43}$$

and in dimensional quantities,

$$\left|\dfrac{d[ES]}{dt}\right| \ll k_a[E]_0([S]+[ES]). \tag{44}$$

Relation (42) is the general condition that defines the time interval where a simplified equation without the first ("transient") term holds. This interval, which contains the $\dfrac{d^2\sigma}{d\tau^2}$ zero-crossing time, may be broad or narrow, depending on the parameters, see Fig. 2.

Eq. (42) encompasses both the QSSA and the pre-equilibrium approximation. The mathematical formulation of the quasi-steady state approximation (QSSA), $\dfrac{d[ES]}{dt}=0$, is too drastic. The condition $\dfrac{d^2\sigma}{d\tau^2}=0$ is misleading, and cannot be obeyed as such. Taken at face value, it implies that $\sigma$ and $p$ evolve linearly with time, that is, according to zero order kinetics. This is not the general behavior.

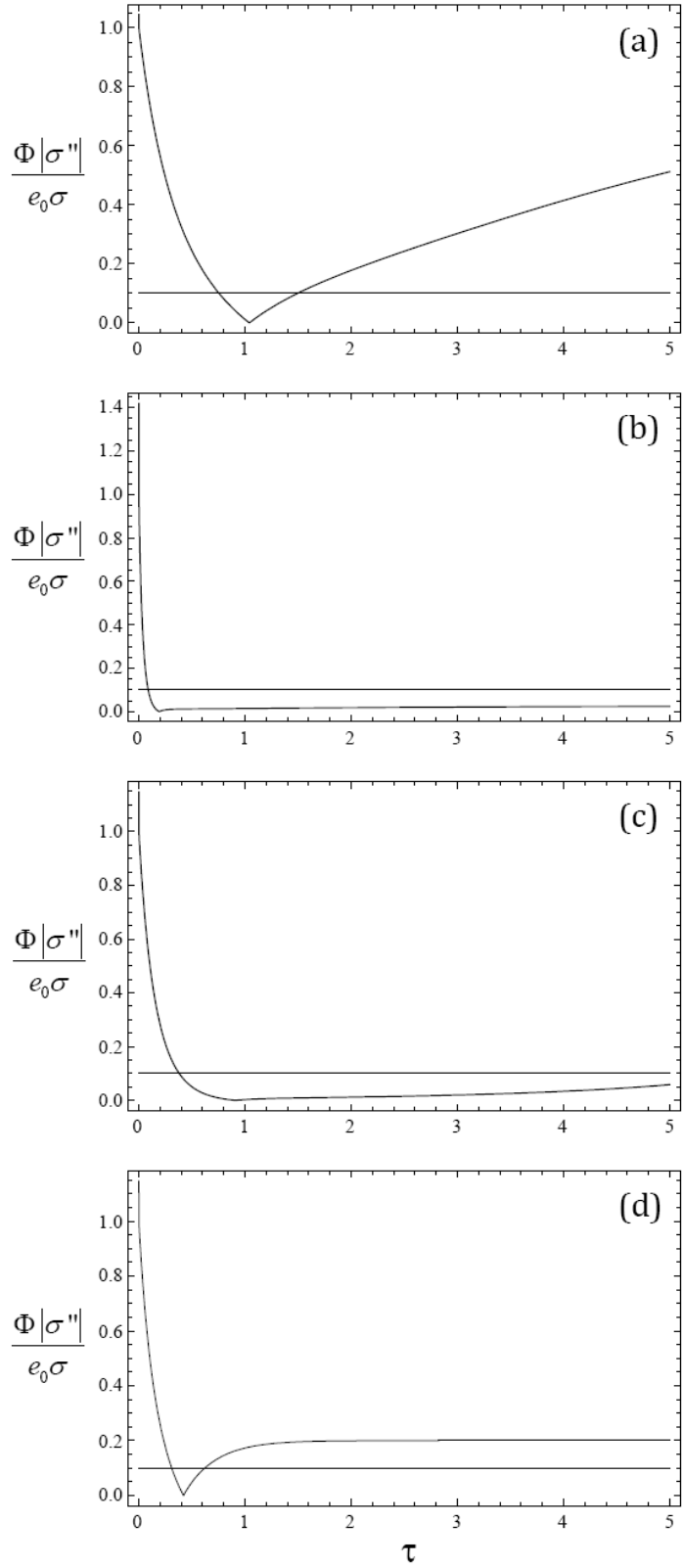

**Fig. 2** Plot of $\dfrac{\Phi}{e_0\sigma}\left|\dfrac{d^2\sigma}{d\tau^2}\right|$ as a function of the reduced time $\tau$ for $e_0 = s_0 = \Phi = 1$ (a), $e_0 = s_0 = 1$, and $\Phi = 0.1$ (b), $e_0 = \Phi = 1$, and $s_0 = 5$ (c), and $e_0 = 5$, and $s_0 = \Phi = 1$ (d). The horizontal line is arbitrarily set at 0.1, and defines the time interval within which $\dfrac{\Phi}{e_0\sigma}\left|\dfrac{d^2\sigma}{d\tau^2}\right| < 0.1$. The respective characteristic times $\tau^*$ [Eq. (46)] are: 0.33 (a), 0.033 (b), and 0.14 (c, d).

The condition $\frac{d^2\sigma}{d\tau^2}=0$ is true only for a specific time, and a linear variation of $p$ and $\sigma$ with time cannot be correct, not even approximately, for the early and late parts of the reaction. Moreover, under certain conditions where Eq. (42) still holds, $\frac{d^2\sigma}{d\tau^2}\simeq 0$ may not be fulfilled at all in any interval of sufficient width.

Given the zero time value of the second derivative of $\sigma$, Eq. (42) cannot hold for very short times. The Maclaurin expansion of $\frac{d^2\sigma}{d\tau^2}$ is:

$$\frac{d^2\sigma}{d\tau^2}=-\frac{e_0 s_0}{\Phi}+\frac{1}{\Phi^2}e_0 s_0\left(1+e_0+s_0\right)\tau-\ldots\simeq -\frac{e_0 s_0}{\Phi}\left[1-\frac{1}{\Phi}\left(1+e_0+s_0\right)\tau\right], \qquad (45)$$

and therefore condition (42) should be fulfilled for reduced times longer than

$$\tau^* \approx \frac{\Phi}{1+e_0+s_0}. \qquad (46)$$

This time defines (approximately) the end of the so-called transient phase, see Fig. 2. Note however that the interval of validity of Eq. (42) may be finite, and thus a second "transient" region can occur for late times, see Fig. 2.

The average cycle time, Eq. (17), has in reduced variables the following form, $\tau_c=\Phi\left(1+\frac{1}{s}\right)$ and therefore starts at $\tau_c(0)=\Phi\left(1+\frac{1}{s_0}\right)$ and increases progressively to infinity. $\tau_c(0)$ is always much larger than $\tau^*$.

For the minimum reduced time $\tau^*$, and using the linear truncated form of Eq. (38),

$$\frac{s(\tau^*)}{s_0}\simeq \frac{1+s_0}{1+e_0+s_0}. \qquad (47)$$

In this way, for $s$ to be still essentially undepleted at the end of the transient period, one must have $e_0 \ll 1+s_0$, which means that for $s_0 \gg 1$ the condition is $e_0 \ll s_0$, and that for $s_0 \ll 1$ the condition is $e_0 \ll 1$ irrespective of the value of $s_0$. This is precisely the condition obtained by Segel for the "complex (fast) time scale".[4]

In the case of pre-equilibrium ($\Phi \ll 1$), there is an almost instantaneous drop in $s$, and a corresponding rise in $c$ owing to pre-equilibration, after which $s$ and $c$ decay very slowly ($k_{cat} \ll k_d$) owing to product formation, see Fig. 3.

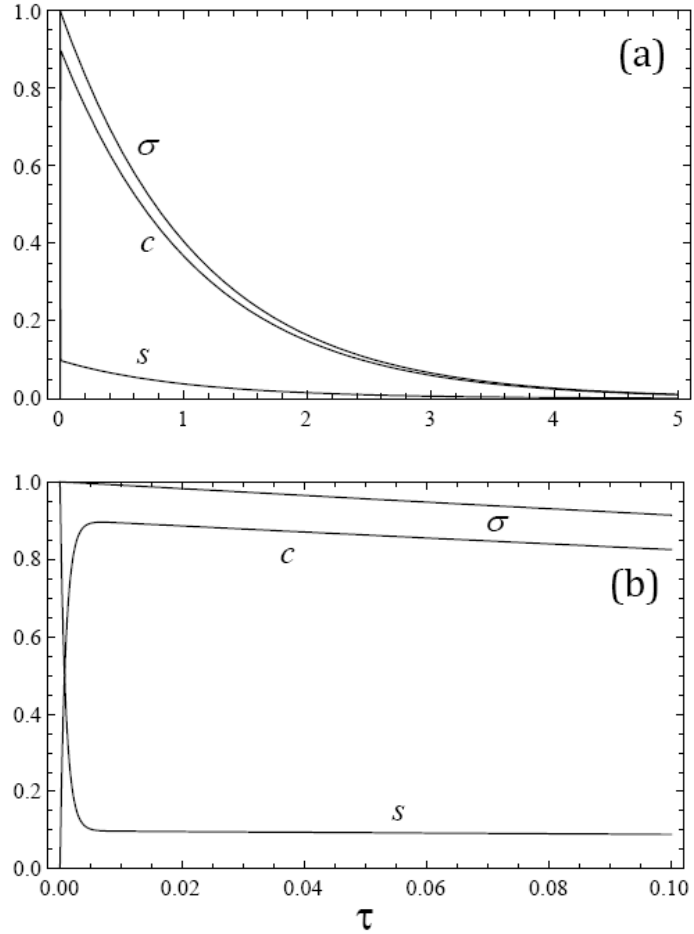

**Fig. 3** Plot of $\sigma$, $s$ and $c$ as a function of the reduced time $\tau$ in a situation of pre-equilibrium: $\Phi$ = 0.01, $e_0$ = 10, and $s_0$ = 1.

These changes occur within the transient period and are not observed for $\sigma$ as $s + c$ remains constant. In this way, if $\Phi \ll 1$ the transient period is very short and $\sigma(\tau^*) = s_0$ irrespective of the values of $s_0$ and $e_0$.

When condition (42) applies, Eq. (31) reduces to

$$\left(1 + e_0 + \sigma + \frac{d\sigma}{d\tau}\right)\frac{d\sigma}{d\tau} + e_0\sigma = 0. \qquad (48)$$

This is a quadratic equation in $\frac{d\sigma}{d\tau}$ that gives (only one of the roots ensures that $\frac{d\sigma}{d\tau} \to 0$ as $\tau \to \infty$)

$$-\frac{d\sigma}{d\tau} = \frac{2e_0\sigma}{1+e_0+\sigma+\sqrt{(1+e_0+\sigma)^2 - 4e_0\sigma}}. \tag{49}$$

This equation was obtained before in dimensional form, see refs. 8,9 and references therein, but does not seem to have been fully explored. Eq. (49) is an expression for the reduced concentration of the substrate-enzyme complex, $c$, cf. Eq. (34). Given that $\sigma = c + s$, Eq. (49) allows to obtain a relation between $c$ and $s$,

$$c = \frac{2e_0 s}{1+s+\sqrt{(1+s)^2 + 8e_0 s}}. \tag{50}$$

This equation is approximate (except for the instant when $c$ attains its maximal value) and holds only after the transient period. For $s = s_0$ it gives

$$c_0 = \frac{2e_0 s_0}{1+s_0+\sqrt{(1+s_0)^2 + 8e_0 s_0}}. \tag{51}$$

The correct result is $c_0 = 0$. In this way, imposing that $c_0 \ll s_0$, one gets from Eq. (51) that $e_0 \ll 1 + 3s_0$, which is close to the condition derived before, $e_0 \ll 1 + s_0$. This ensures that the transient period is short compared to the reaction time, i.e., that the two phase plane $c = c(s_0)$ curves, one exact and the other approximate, have an early crossing near $s_0$.

Eq. (49) is free from approximations other than that it holds only after the transient period. It can thus be integrated within its range of validity. Writing $\sigma(\tau^*) = \sigma^*$, where $\tau^*$ is given by Eq. (46), it is obtained that

$$(1+e_0)\ln\left(\frac{\sigma^*}{\sigma}\right) + (\sigma^* - \sigma) + \int_\sigma^{\sigma^*} \sqrt{\left(\frac{1+e_0}{u}\right)^2 + 2\left(\frac{1-e_0}{u}\right) + 1}\, du = 2e_0(\tau - \tau^*). \tag{52}$$

The integral in Eq. (52) is analytical but the explicit form is too cumbersome, and for this reason the above representation is preferable. The only point where this equation holds exactly is where the enzyme-substrate complex concentration attains a maximum.

Eq. (52) is valid for $\tau > \tau^*$. In order to cover essentially the whole time domain, $s$ should be close to $s_0$, unless $s$ decays very fast during a short transient phase (pre-equilibrium). The sufficient condition for $s(\tau^*) \simeq s_0$ was obtained from Eq. (47), and is $e_0 \ll 1+ s_0$. In such a case $\tau^*$ is close to zero. The condition $e_0 \ll 1+\sigma < 1+s_0$ implies that Eq. (49) simplifies to

$$\frac{d\sigma}{d\tau} = -\frac{e_0 \sigma}{1+\sigma}, \qquad (53)$$

hence

$$\ln\left(\frac{s_0}{\sigma}\right) + (s_0 - \sigma) = e_0 \tau. \qquad (54)$$

This can be rewritten as

$$\sigma = W\left(s_0 e^{s_0} e^{-e_0 \tau}\right), \qquad (55)$$

where $W$ is the Lambert function. In this way, the dimensionless substrate concentration becomes

$$s = W\left(s_0 e^{s_0} e^{-e_0 \tau}\right) - e_0 \frac{W\left(s_0 e^{s_0} e^{-e_0 \tau}\right)}{1+W\left(s_0 e^{s_0} e^{-e_0 \tau}\right)}, \qquad (56)$$

and reduces to[7] $W\left(s_0 e^{s_0} e^{-e_0 \tau}\right)$ only when $e_0 \ll 1+\sigma$. It would seem that Eqs. (55) and (56) are equivalent, as follows from the approximation used to obtain Eq. (49), however Eq. (56) is a much better approximation than Eq. (55) when $e_0$ is lower than $1+s_0$ by just one order of magnitude, see Fig. 4.

Eq. (54) can be used to define a characteristic reaction time. According to common usage, $t_{1/2}$ is the time necessary for the reactant concentration to attain half of the initial value. Given the meaning of $\sigma$, and the nature of the HMM mechanism, which has an intermediate species, it is preferable to define such a parameter as the time necessary for the concentration of product to reach half of the final value. In this way, for $\sigma = s_0/2$ the characteristic time is

$$\tau_{1/2} = \frac{\ln 2 + s_0}{2e_0}. \qquad (57)$$

This value is close to the "substrate (slow) time scale" defined by Segel,[4] but has a precise meaning.

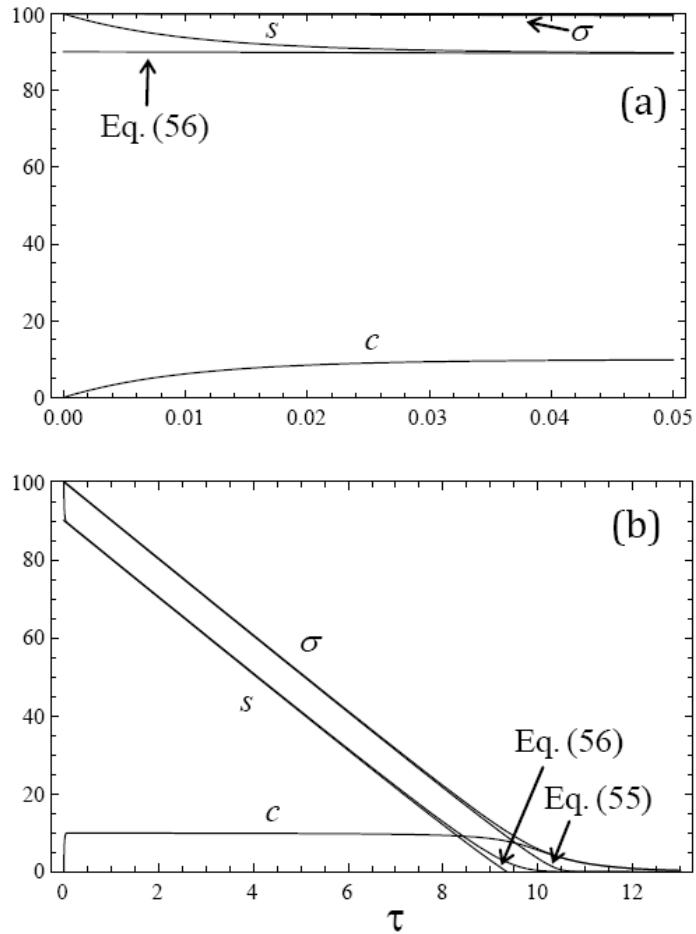

**Fig. 4** Plot of $\sigma$, $s$ and $c$ as a function of the reduced time $\tau$ for $e_0 = 10$, $s_0 = 100$, and $\Phi = 1$. Also shown are the curves computed with Eqs. (55) and (56).

For $s_0 \ll 1$ (implying $e_0 \ll 1$) the characteristic reaction time is independent from $s_0$, whereas for $s_0 \gg 1$ (implying $e_0 \ll s_0$) it is a function of the ratio $s_0/e_0$, which is always higher than 1. In this last situation, the characteristic reaction time is proportional to $s_0$ and inversely proportional to $e_0$. Using Eq. (20) with $[P] = [S]_0/2$ and a cycle duration of $1/(k_d+k_{cat})$, the same result is obtained for $t_{1/2}$. In terms of the dimensional parameters Eq. (57) reads,

$$t_{1/2} = \frac{[S]_0 + K_m \ln 2}{2k_{cat}[E]_0} = \frac{[S]_0}{2k_{cat}[E]_0} + \frac{\ln 2}{2\Phi k_a[E]_0}. \tag{58}$$

The determination of $t_{1/2}$ as a function of the initial concentrations is a simple procedure to obtain $k_{cat}$ and $K_m$. For fixed initial enzyme concentration, for instance, a plot of $t_{1/2}$ vs. $[S]_0$ is linear, and $k_{cat}$ and $K_m$ are obtained from the slope and intercept. Note that the above analysis is not limited to $t_{1/2}$. A characteristic time $t_{1/n}$ can also be used, with $n = 3, 4$, etc., allowing for instance to use measurements made at earlier reaction times, where product inhibition is negligible.

In the case of pre-equilibrium ($\Phi << 1$) $\tau^* \approx 0$ and Eq. (52) with $\sigma^* = s_0$ applies from the beginning, irrespective of $e_0$ and $s_0$,

$$(1+e_0)\ln\left(\frac{s_0}{\sigma}\right) + (s_0 - \sigma) + \int_\sigma^{s_0} \sqrt{\left(\frac{1+e_0}{u}\right)^2 + 2\left(\frac{1-e_0}{u}\right) + 1}\, du = 2e_0\tau. \tag{59}$$

A characteristic reaction time $\tau_{1/2}$ similar to Eq. (57) can be obtained from this equation,

$$\tau_{1/2} = \frac{1+e_0}{2e_0}\ln 2 + \frac{s_0}{4e_0} + \frac{1}{2e_0}\int_{1/2}^{1} \sqrt{\left(\frac{1+e_0}{x}\right)^2 + 2\left(\frac{1-e_0}{x}\right)s_0 + s_0^2}\, dx. \tag{60}$$

This characteristic time is displayed in Fig. 5 as a function of $s_0$ and for several values of $e_0$.

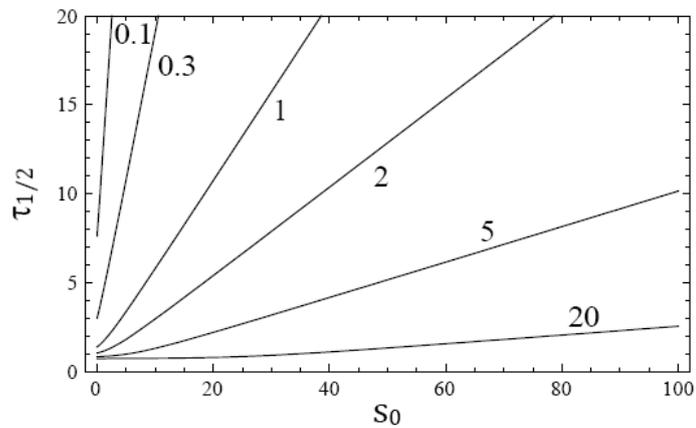

**Fig. 5** Plot of $\tau_{1/2}$ [Eq. (60)] as a function of $s_0$. The number next to each curve is the respective $e_0$.

For $s_0 \ll 1+e_0$, Eq. (60) reduces to

$$\tau_{1/2} = \left(1 + \frac{1}{e_0}\right)\ln 2 + \frac{s_0}{4e_0}, \qquad (61)$$

see Fig. 5, while for $s_0 \gg 1+e_0$, Eq. (60) becomes

$$\tau_{1/2} = \frac{s_0}{2e_0}, \qquad (62)$$

see Fig. 5. For $e_0 \ll 1$ Eq. (57) is recovered from Eq. (60), whereas for $e_0 \gg 1$ Eq. (60) becomes

$$\tau_{1/2} = \ln 2 + \frac{s_0}{2e_0}. \qquad (63)$$

When $e_0 \ll 1+s_0$ the integrated form of Eq. (59) reduces to Eq. (56). When $e_0 \gg 1$ Eq. (59) becomes

$$(1+e_0)\ln\left(\frac{s_0}{\sigma}\right) + (s_0 - \sigma) + \int_\sigma^{s_0}\left|\frac{e_0}{u} - 1\right|du = 2e_0\tau. \qquad (64)$$

If $s_0 < e_0$, the integrated form is

$$\sigma = s_0 e^{-\tau}. \qquad (65)$$

If $s_0 > e_0$, one has (Appendix B)

$$\sigma = \begin{cases} s_0 - e_0\tau & \text{if } \tau < \dfrac{s_0}{e_0} - 1, \\[6pt] e_0 e^{\frac{s_0}{e_0}-1} e^{-\tau} & \text{if } \tau > \dfrac{s_0}{e_0} - 1. \end{cases} \qquad (66)$$

hence the decay of $\sigma$ is initially linear but changes abruptly to an exponential when $\sigma = e_0$, i.e., switches instantaneously from zero-order to first-order kinetics. The corresponding evolutions for $s$, $c$ and $p$ can be obtained from Eqs. (32), (34) and (40). A curious result is obtained for $s$: it is always zero in the case $s_0 < e_0$, and in the case $s_0 > e_0$ decays linearly to zero, becoming zero for $\tau = \dfrac{s_0}{e_0} - 1$. The corresponding results for $c$ are an exponential decay in the first case ($c = \sigma$), and a time-independent profile ($c = e_0$), followed by an exponential decay ($c = \sigma$, after $\tau = \dfrac{s_0}{e_0} - 1$) in the second. These results are not exact, and somewhat smoother transitions between the regimes are in fact observed, see Fig. 6.

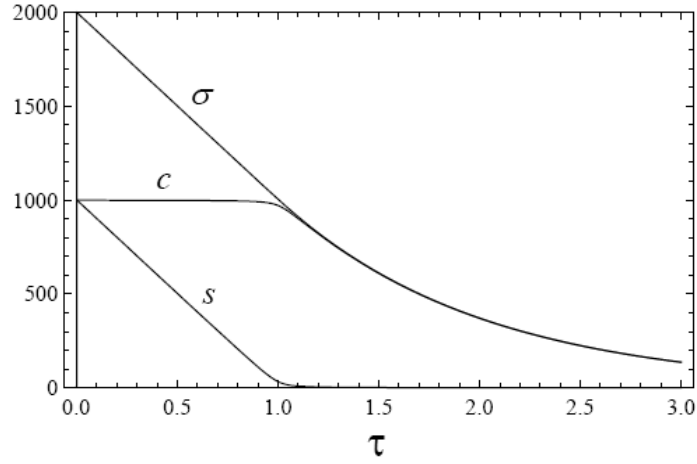

**Fig. 6** Plot of σ, *s* and *c* as a function of the reduced time τ for $e_0$ = 1000, $s_0$ = 2000, and Φ = 0.01. The behavior for Φ = 1 is very similar.

**VI.2.2 Excess enzyme and related cases**

When 1+$e_0$ >> $s_0$ (which means that for $e_0$ >> 1 one has $e_0$ >> $s_0$ for any $s_0$, and that for $e_0$ << 1 one has $s_0$ << 1, $e_0$ being smaller or larger than $s_0$), Eq. (31) becomes linear

$$\Phi \frac{d^2\sigma}{d\tau^2} + (1+e_0)\frac{d\sigma}{d\tau} + e_0 \sigma = 0, \qquad (67)$$

and has an exact solution, that covers both the initial, transient phase as well as the subsequent phase:

$$\sigma(\tau) = \frac{s_0}{\lambda_2 - \lambda_1}\left(\lambda_2 e^{-\lambda_1 \tau} - \lambda_1 e^{-\lambda_2 \tau}\right), \qquad (68)$$

with

$$\lambda_{1,2} = \frac{1+e_0 \pm \sqrt{(1+e_0)^2 - 4e_0\Phi}}{2\Phi}. \qquad (69)$$

The short component has a lifetime

$$\frac{1}{\lambda_1} = \frac{2\Phi}{1+e_0+\sqrt{(1+e_0)^2-4e_0\Phi}}, \tag{70}$$

while the long component is

$$\frac{1}{\lambda_2} = \frac{1+e_0+\sqrt{(1+e_0)^2-4e_0\Phi}}{2e_0}. \tag{71}$$

For a given $e_0$, the difference between the two lifetimes is the larger the smaller the $\Phi$, i.e., the duration of the transient phase is again affected by the enzyme efficiency, see Figs. 3 and 7.

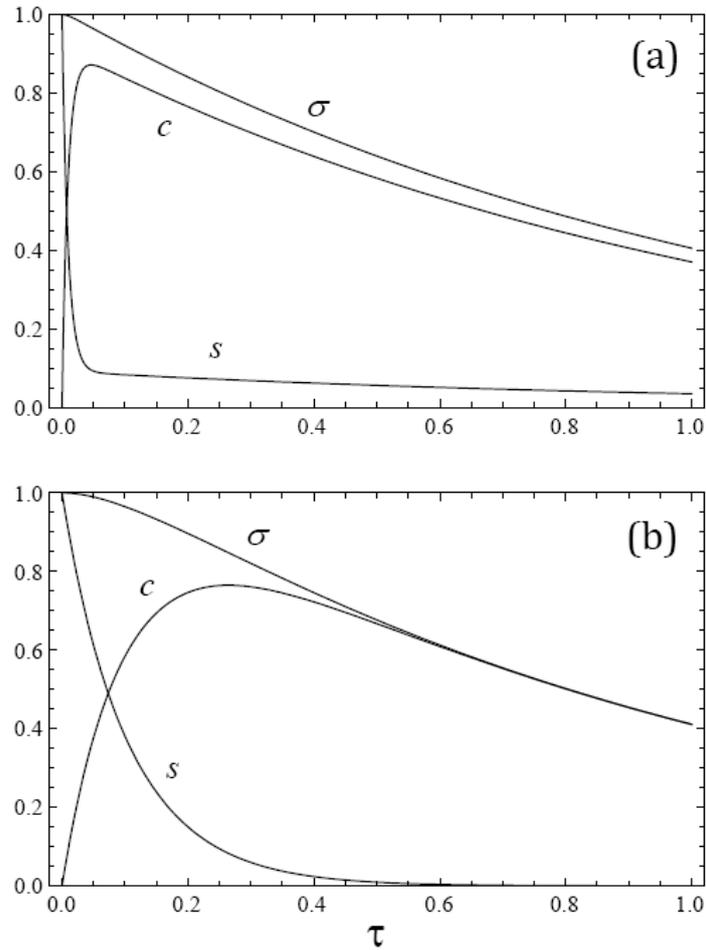

**Fig. 7** Plot of $\sigma$, $s$ and $c$ as a function of the reduced time $\tau$ for $e_0$ = 10, $s_0$ = 1, with $\Phi$ =0. 1 (a) and $\Phi$ = 1 (b).

For $\Phi \ll 1$ Eq. (70) reduces to $\frac{1}{\lambda_1} = \frac{\Phi}{1+e_0}$, and Eq. (71) to $\frac{1}{\lambda_2} = \frac{1+e_0}{e_0}$. For $e_0 \gg 1$, $\frac{1}{\lambda_2} = 1$.

## VI.2.3 The initial transient phase

During the initial phase of the reaction, if the substrate $s$ is still essentially undepleted, Eq. (31) can be simplified to

$$\Phi \frac{d^2\sigma}{d\tau^2} + (1+e_0+s_0)\frac{d\sigma}{d\tau} + e_0 s_0 = 0, \qquad (72)$$

by assuming that $s \simeq s_0$. This is again a linear differential equation with an exact solution,

$$\sigma = s_0 \left[ \frac{e_0 \Phi}{(1+e_0+s_0)^2}\left(1-\exp\left[-\frac{(1+e_0+s_0)\tau}{\Phi}\right]\right) + \frac{1+e_0(1-\tau)+s_0}{1+e_0+s_0} \right]. \qquad (73)$$

This solution encompasses the usual one, Eq. (29), which reads in dimensionless parameters

$$\sigma = s_0 - \frac{e_0 s_0}{1+s_0}\left[\tau - \frac{\Phi}{1+s_0}\left(1-e^{-\frac{1+s_0}{\Phi}\tau}\right)\right]. \qquad (74)$$

In fact, the first corresponds to a time evolution of the enzyme-substrate complex given by

$$c = \frac{e_0 s_0}{1+e_0+s_0}\left(1-e^{-\frac{1+e_0+s_0}{\Phi}\tau}\right), \qquad (75)$$

whereas the second is

$$c = \frac{e_0 s_0}{1+s_0}\left(1-e^{-\frac{1+s_0}{\Phi}\tau}\right). \qquad (76)$$

It is thus seen that they coincide only for $e_0 \ll 1+s_0$.

A different approximation for the transient phase is to assume that it is the total substrate $\sigma$ that is essentially undepleted, $\sigma = s_0$, as happens in pre-equilibrium conditions, and Eq. (31) becomes

$$\Phi \frac{d^2\sigma}{d\tau^2} + \left(1+e_0+s_0+\frac{d\sigma}{d\tau}\right)\frac{d\sigma}{d\tau} + e_0 s_0 = 0. \qquad (77)$$

The approximation corresponds to the simplified scheme $E+S \underset{k_d}{\overset{k_a}{\rightleftarrows}} ES$. This leads to a more complicated but in some cases more accurate solution, as it may cover a slightly larger time interval (some *s* is allowed to change into *c*, but not into *p*). The solution is obtained by changing Eq. (77) into

$$\Phi \frac{dc}{d\tau} + (1 + e_0 + s_0 - c)c - e_0 s_0 = 0. \tag{78}$$

It is obtained that

$$c = \frac{1}{2}\left[\alpha - \gamma \frac{(\alpha + \gamma)^2 + \beta \exp\left(-\frac{\gamma}{\Phi}\tau\right)}{(\alpha + \gamma)^2 - \beta \exp\left(-\frac{\gamma}{\Phi}\tau\right)}\right], \tag{79}$$

with $\alpha = 1 + e_0 + s_0$, $\beta = 4 e_0 s_0$, and $\gamma = \sqrt{\alpha^2 - \beta}$. The concentration starts from zero, displays a rise-time, and then stabilizes at the initial equilibrium value. From this equation, all other concentrations can be computed for the transient phase.

## VII. CONCLUSIONS

A different and more general view of Henri-Michaelis-Menten enzyme kinetics was presented.

In the first part of the paper, a simplified but useful description that stresses the cyclic nature of the catalytic process was introduced. This approach provides parameters that define enzyme efficiency [product yield per cycle, Eq. (15)] and speed [average cycle duration, Eq. (17)]. The turnover rate can be described in terms of these two parameters by an equation [Eq. (18)] that has a clear dynamical meaning. The total number of cycles per enzyme was also obtained [Eq. (19)]. The time-dependence of the substrate concentration after the initial transient phase [Eq. (22)] is derived in a simple way that dispenses the mathematical technique known as quasi-steady-state approximation.

In the second part of the paper (Sect. VI), an exact one-dimensional formulation of HMM kinetics was obtained. The whole problem is condensed in a single one-variable evolution equation, Eq. (31), which is a second-order non-linear differential equation, and the control parameters are reduced to just three dimensionless quantities: the enzyme efficiency $\Phi$, the substrate reduced initial concentration, $s_0$, and the enzyme reduced initial concentration, $e_0$. The exact solution of HMM kinetics was obtained from Eq. (31) as a set of Maclaurin series. From the same equation, a number of approximate solutions, some known, some new, were derived in a systematic way that allowed a precise evaluation of the respective level of approximation and conditions of validity. The role of pre-equilibrium and quasi-steady-state approximations was in particular defined in a more satisfactory way. Eqs. (56), (57), (59), and (60) are significant new results. Eq. (31) was also shown to be adequate for the numerical computation of the concentrations of all species as a function of time for any given combination of parameters.

# APPENDIX A

Eq. (21) can be integrated by separation of variables to give

$$\ln\frac{[S]_0-[P]}{[S]_0} - \frac{[P]}{K_m} = -\frac{k_{cat}[E]_0}{K_m}t. \tag{A1}$$

Or, using $[S]=[S]_0-[P]$,

$$\ln\frac{[S]}{[S]_0} + \frac{[S]-[S]_0}{K_m} = -\frac{k_{cat}[E]_0}{K_m}t. \tag{A2}$$

This integrated form has been known for a long time (see ref. 14 and ref. therein). More recently, and with the help of technical software, Schnell and Mendoza[7] recognized that the substrate concentration can be explicitly given as a function of time in terms of the Lambert $W$ function. This function is defined as the inverse function of $x\,e^x$, that is, $W(x)\,e^{W(x)} = x$. Indeed, Eq. (A2) can be rewritten in the form

$$\left(\frac{[S]}{K_m}\right)\exp\left(\frac{[S]}{K_m}\right) = \left(\frac{[S]_0}{K_m}\right)\exp\left(\frac{[S]_0}{K_m}\right)\exp\left(-\frac{k_{cat}[E]_0}{K_m}t\right), \tag{A3}$$

hence[7]

$$[S](t) = K_m W\left[\left(\frac{[S]_0}{K_m}\right)\exp\left(\frac{[S]_0}{K_m}\right)\exp\left(-\frac{k_{cat}[E]_0}{K_m}t\right)\right]. \tag{A4}$$

The physical content of Eq. (A4) is of course identical to that of Eq. (A2), but the explicit form is more elegant and also more convenient for computational purposes, as the function is implemented in technical computing software (e.g. as *ProductLog*[*x*] in *Mathematica*).

Given that for small $x$ the Lambert function becomes $W(x) = x$, and that for large $x$ the function is $W(x) \approx \ln x$, Eqs. (26) and (25) follow from Eq. (A4).

## APPENDIX B

For $s_0 > e_0$ and $\sigma > e_0$, Eq. (60) becomes

$$(1+e_0)\ln\left(\frac{s_0}{\sigma}\right) + (s_0 - \sigma) - \int_\sigma^{s_0}\left(\frac{e_0}{u} - 1\right)du = 2e_0\tau, \tag{B1}$$

or

$$\sigma = s_0 - e_0\tau, \tag{B2}$$

while for $s_0 > e_0$ and $\sigma < e_0$ Eq. (60) becomes

$$(1+e_0)\ln\left(\frac{s_0}{\sigma}\right) + (s_0 - \sigma) + \int_\sigma^{e_0}\left(\frac{e_0}{u} - 1\right)du - \int_{e_0}^{s_0}\left(\frac{e_0}{u} - 1\right)du = 2e_0\tau, \tag{B3}$$

or

$$\sigma = e_0\, e^{\frac{s_0}{e_0} - 1}\, e^{-e_0\tau}. \tag{B4}$$

## ACKNOWLEDGMENTS

This work was supported by Fundação para a Ciência e a Tecnologia (FCT, Portugal) and POCI 2010 (FEDER) within project PTDC/ENR/64909/2006.